\documentclass[nobib]{tufte-handout}
\usepackage[backend=bibtex, natbib=true,maxbibnames=99]{biblatex}
\usepackage{hyphenat}
\usepackage[utf8]{inputenc}
\usepackage{eurosym}
\usepackage[utf8]{inputenc}
\DeclareUnicodeCharacter{20AC}{\euro}
\usepackage[compatibility=false]{caption} %Otherwise \autoref dos not work with tufte
\usepackage{hyperref}
\usepackage{graphicx}
\usepackage{amsmath}
\usepackage{amssymb}
\usepackage{listings}
\usepackage{xcolor}
\colorlet{punct}{red!60!black}
\definecolor{background}{HTML}{EEEEEE}
\definecolor{delim}{RGB}{20,105,176}
\colorlet{numb}{magenta!60!black}

\lstdefinelanguage{json}{
    basicstyle=\normalfont\ttfamily,
    numbers=left,
    numberstyle=\scriptsize,
    stepnumber=1,
    numbersep=8pt,
    showstringspaces=false,
    breaklines=true,
    frame=lines,
    backgroundcolor=\color{background},
    literate=
     *{0}{{{\color{numb}0}}}{1}
      {1}{{{\color{numb}1}}}{1}
      {2}{{{\color{numb}2}}}{1}
      {3}{{{\color{numb}3}}}{1}
      {4}{{{\color{numb}4}}}{1}
      {5}{{{\color{numb}5}}}{1}
      {6}{{{\color{numb}6}}}{1}
      {7}{{{\color{numb}7}}}{1}
      {8}{{{\color{numb}8}}}{1}
      {9}{{{\color{numb}9}}}{1}
      {:}{{{\color{punct}{:}}}}{1}
      {,}{{{\color{punct}{,}}}}{1}
      {\{}{{{\color{delim}{\{}}}}{1}
      {\}}{{{\color{delim}{\}}}}}{1}
      {[}{{{\color{delim}{[}}}}{1}
      {]}{{{\color{delim}{]}}}}{1},
}

\title{Easy steps towards a sane IT policy in hospitals}
\author{Edouard Klein}
\date{\today}

\begin{document}

\maketitle

\begin{abstract}
We witnessed the low quality of IT solutions in Paris hospitals. The price paid to private companies for these solutions and the cost incurred from their inefficiency constitute a gross and appalling waste of public resources. We propose to bootstrap a change in IT policy by having heads of department hire IT workers ; we give advice to the central decision making body on how to incentivize them. Easily measurable efficiency gains as well as hard-to-quantify positive externalities will follow.
\end{abstract}

\setcounter{tocdepth}{4}
\tableofcontents

\section{Introduction}
\label{sec:introduction}

In Paris hospitals, there is no easy way to access medical records although those are computerized since circa 2011.

In order to gather data prior to statistical analysis, one has to tackle the task of manually searching and extracting this data from a software called Cristal-Net.

This software is technologically outdated (it can only run on Windows XP with IE 6\footnote{Those pieces of software first came out in 2001 (more than 13 years ago at the time of writing), and support from Microsoft was finally dropped in April 2014.}). It is also slow, prone to crashing, not devoid of  bugs and has a terrible user experience. On top of that, it has been very costly to build and is, I believe, costly to maintain. Cristal-Net being presented as modern and efficient \cite{cristalnet}, it is my understanding that most medical record management systems are as bad or even worse.

The boring and error-prone task of manual data entry is usually devolved to those at the bottom of the hospital/med-school food-chain : the medical residents.

Those brave soon-to-be-doctors, outrageously overworked as they are \cite[in french]{internesFIXME}, have to spend countless hours on this desk-drone task, despite the fact that they have been studying medicine for close to or more than ten years.

This is not a problem of pride at all. Our point is not to say that medical residents are somehow too good for lowly tasks like mindless data entry. They themselves do not pipe up and do this job with application.

Our point is that there are two grave, disturbing, compounding problems here. The first is that the skilled manpower directly lost in data entry tasks could be used elsewhere where it is scarce and useful : in the operating room, talking with and listening to the patients, in staff reviews or in the classroom (or catching up on sleep or social life).

The second is that medical residents are for the vast majority greatly under-skilled in all things related to Information Technology (IT). Because of that, they will be quite inefficient at their data-entry tasks. They may pollute the data by making entry mistakes or make errors that will lead to data loss, data corruption or loss of patient confidentiality. Datasets that are actually easy to create will never see the light of day because the IT-unskilled operators will see the task as too daunting or downright impossible.

The cost of managing the mistakes of the (IT-)unskilled operators, and the cost of opportunity of never seeing those datasets that-could-have-been come in addition  to the direct loss of manpower and money that the creation, use and maintenance of low-quality software such as Cristal-Net entails.

In this context of funds and manpower shortage in pubic hospitals in France, it is a shame to witness such a gross misallocation of public resources.

We propose a two-fold solution to this problem. We first give a technical emergency fix in the form a small piece of software.
\begin{itemize}
\item This software has successfully been used to reliably extract specific variables from more than a thousand medical records accessible only through Cristal-Net (\autoref{sec:automatic}).
\item We provide a link to the source code and documentation of this software as well as suggestions to hospital managers and heads of department regarding its deployment and adaptation to any record management software (not just Cristal-Net) (\autoref{sec:deployment}).
\item We provide a rationale for the quick and generalized hiring of IT workers in any hospital where there is no solution for automated data gathering from medical records (\autoref{sec:economics}).
\end{itemize}

The second part of the solution is political. 
\begin{itemize}
\item We propose our vision of what a well-thought out medical record management system would look like (\autoref{sec:uptodate}).
\item We explain how such a solution will not come from a central decision making body but from interaction, discussion and collaboration between IT field workers (\autoref{sec:good}).
\item We finally advise higher-ups about what kind of incentives need to be put in place for this to happen, and also what kind of schemes ought \emph{not} to be put in place in order to avoid perverse incentives (\autoref{sec:incentives}).
\end{itemize}

We provide these ideas to sparkle discussion about software and IT policies in public health institutions and to pave the way to a saner management of skilled human resources and public funds. We show these ideas align with what happen in the field of software development and with what is called for by members of the medical community with respect to the computerization of the field (\autoref{sec:related}).

\section{The data-entry task}
\label{sec:dataentry}

\subsection{Manual data-entry}
\label{sec:manual}

The task at hand is quite straightforward. We have a list of patients (identified by a unique number) that have given birth with a planned C-section, and we want to know the hemoglobin rate well before the C-section (base rate) and just after. This data will be used as part of a comparative study between oxytocin and carbetocin. All the women before a certain date were injected with oxytocin, and then the protocol changed and all the women were injected with carbetocin instead. There is no doubt this data will deliver beautiful insights about the respective risks and benefits of the two products involved as there probably are no confounding factors : the change of protocol was well-defined in time and systematic, and quite a lot of data has been gathered (in the hundreds of patients in each group).

The two points of data we want exist in a database somewhere, but the digital medical record management system of Paris hospitals make them available only through Cristal-Net.

To access data for one patient, in the best case\footnote{If Cristal-Net does not stop working, if the medical records are complete and if the operator does not make any mistake.}, a minimum of 41 actions (clicks, or hotkey presses) are required. The constant lagging of Cristal-Net prevents the operator from mindlessly executing these operations as one would play a tune one knows well on a piano, for example. The operator has to wait sometimes up to 15 or 20 seconds for feedback from the interface before he can proceed with the next action. The video that we use to describe the software (see \autoref{sec:automatic}) gives a good idea of the tediousness of this task.

Three hours of manual work yielded data for around 80 patients, that is close to two minutes and fifteen seconds per record. Over 1100 records had to be analyzed, which made for an expected workload of over 41 man-hours. 

\subsection{Automatic data-entry}
\label{sec:automatic}

We were able, in less than a third of that time, to create a piece of software automating this process. This software contains very little actual algorithmic logic\footnote{We only need to compare the dates of the records to the date of the C-section.}. The grunt of its work is to reproduce the various clicks and keypresses the operator would make, wait for visual feedback from Cristal-Net, and gracefully handle edge-cases and errors. The software really, actually takes the place of the human operator : it simulates mouse and keyboard events and analyses images on the screen to make decisions\footnote{Yeah, data entry is so dumb a machine can do it.}.

A video of this software in action can be found at \url{https://www.youtube.com/watch?v=nGMSP_wBr-8}. Bear in mind that all throughout the video, there is no human input, everything is automated.

The software runs with little human supervision. If everything fails, it can be relaunched back from before everything went awry. An (IT-)unskilled medical resident was trained in its use in less than half an hour and was able to supervise its execution, analyze and report problems, and relaunch the system until completion of the data-gathering task. Supervising the execution of the program does not need to be done in real time. One can go have a look once in a while, and, if everything is not running smoothly, purge the error logs and relaunch the thing.

Now that we overcame the various specific problems of creating this software for the first time, we are fairly confident in our ability (or any mid-level software engineer's ability) to adapt it to a new data gathering task in, say, two to six hours.

\section{The economics of automatic data-entry}
\label{sec:economics}

\subsection{Comparative advantage}
\label{sec:comparative}

The costs-benefits analysis of hiring an IT specialist to do (or automate away) the data-entry jobs instead of using medical residents is quite straightforward. The gain (in monetary units) is the difference between the cost of the medical resident  and the cost of the IT specialist  for the same data-entry task.

\begin{equation}
G = cost(med) - cost(IT)
\end{equation}

We will take the naive approach of thinking that the cost of work is simply the hourly cost $C$ times the time spent $t$. We will see later (\autoref{sec:compounding}) that the benefits of hiring an IT specialist go beyond just him being faster at data-entry than a resident, and that the cost of making a resident do data entry are higher than just paying him. But for now let's just focus on the \emph{comparative advantage} \cite[27-36]{economics} of the IT worker for a data entry task.

\begin{equation}
G = t_{med}\cdot C_{med} - t_{IT}\cdot C_{IT}
\end{equation}

It is economically sound to hire an IT worker if the gain $G$ is positive, or at the very least zero :
\begin{eqnarray}
G =  t_{med}\cdot C_{med} - t_{IT}\cdot C_{IT} &\geq 0\\
 t_{med}\cdot C_{med}  &\geq  t_{IT}\cdot C_{IT}\\
  \frac{t_{med}\cdot C_{med}}{t_{IT}}  &\geq C_{IT}
\end{eqnarray}

We can now talk about the relative (to the resident) efficiency $\alpha$ of the IT worker.
\begin{equation}
\alpha = \frac{t_{med}}{t_{IT}}
\end{equation}
For example if the IT worker can, by working one hour on a task\footnote{Either by doing it himself, or by automating part of it.}, spare two hours of a medical resident's time, then $\alpha=2$. Whether the gain $G$ is positive, and by how much, depends on $\alpha$ :
\begin{equation}
G \geq 0 \Leftrightarrow C_{IT} \leq \alpha C_{med}
\end{equation}

\begin{figure*}
\center
\includegraphics{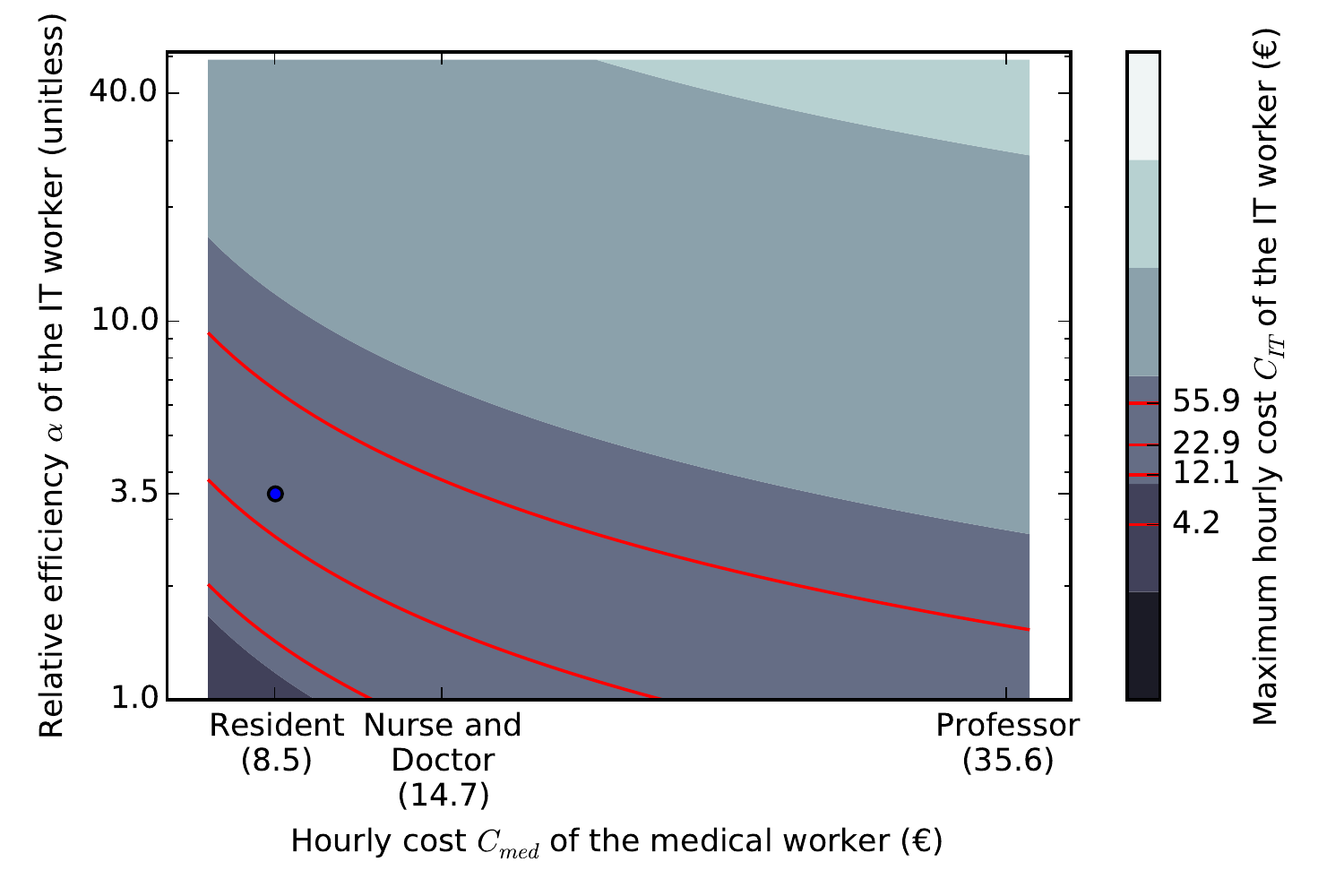}
\caption{Maximum price one should rationally pay to hire an IT worker. Notice the log scale for $\alpha$ and $C_{IT}$. The datapoint at $\alpha=3.5$ and $C_{med} = 8.5$ means we could have been paid an entry level salary (for a cost to the hospital of $22.9$€ per hour) and the hospital still would have benefited from the exchange. Code for plotting this can be found at \url{https://gist.github.com/edouardklein/2d3937eac5c0c4c1b79f} }
\label{fig:IT_cost}
\end{figure*}

A visual representation of this inequality is shown \autoref{fig:IT_cost}. We show $max(C_{IT})$ : how much an IT worker can cost to the hospital and still have the hospital benefit. Ticks for $max(C_{IT})$ are the cost associated with \cite{salaires} :
\begin{itemize}
\item an intern (4.3 €/h),
\item the minimum legal wage (12.1 €/h), 
\item an entry level salary for a software engineer (22.9 €/h),
\item the maximum wage for a software project leader in Paris (55.9 €/h).
\end{itemize}

Ticks for the cost of the medical personnel are located at the approximate actual\footnote{Computed with respect to how many hours these people actually spend working, and not how many they are supposed to spend. The theoretical hourly cost is much higher, especially for residents. Using it would strenghten our argument.\\\vspace{1em}} costs for 
\begin{itemize}
\item residents (8.5 €),
\item nurses and doctors (14.7 €)\footnote{Doctors ("Chef de clinique") are paid twice as much nurses, but work twice as many hours.},
\item professors (35.6 €).
\end{itemize}

Finally, ticks for the relative efficiency $\alpha$ of the IT worker are at :
\begin{itemize}
\item $1$, a theoretical lower bound at which the IT worker is as (in)efficient at data entry as an IT-unskilled worker,
\item  $3.5$ the efficiency we observed in our experiment (\autoref{eq:alpha_true}),
\item $10$, the efficiency we can expect if one tackle a new data entry task using our software (\autoref{sec:compounding}),
\item $40$, the efficiency that the use and deployment of high-quality software in hospitals would allow to attain reliably (\autoref{sec:uptodate}).
\end{itemize}

Thanks to our small program, a task that should have taken a resident 41 hours of work (recall \autoref{sec:manual}) has been done in less than two hours. The time gain for the resident was then $t_{med} = 41-2 = 39$ hours. The time spent writing and testing the program was a bit less than $t_{IT} = 11$ hours. We have, from the start, an efficiency of 
\begin{equation}
\alpha = \frac{t_{med}}{t_{IT}} = \frac{39}{11} \approx 3.5
\label{eq:alpha_true}
\end{equation}

Our experiment shows that from the start, a skilled IT worker will provide enough value to the hospital to compensate for the cost of his employment. This is despite the low cost of the people he is replacing (mainly the residents who cost a whooppingly low 8.5 €\footnote{Combine ridiculously high working hours with a salary computed on theoretically worked hours (instead of actually worked hours) just shy above minimum wage and you've got a salary between half and two third of the legal minimum wage.}) and the difficulty of automating data entry for the first time, as bugs and hoops have to be discovered.

%Therefore, considering the \emph{actual} cost  of a resident-hour to the APHP\footnote{The institution that manages Paris hospitals}, which is close to , hiring an IT worker to automate data entry tasks is worth it up to a hourly cost of $\alpha\cdot C_{med} = 3.5\cdot 6 \approx 21$ €. This is above the cost of a minimum wage worker, and high above the minimal legal retribution for an engineering school intern.

Considering this, it should be a no-brainer to hire a programmer/sysadmin in every unit of every hospital. Or, at the very least, to open positions for engineering schools interns. Yet, the hospital I did this work (for free) in has no dedicated IT support. The one IT correspondent they can talk to is in another hospital, and is tasked with maintenance, not development. This basic, but quantitative analysis shows that there is an urgent need for the APHP to reevaluate its IT policy.

\subsection{Compounding benefit}
\label{sec:compounding}

If we were to tune our program to a new data entry task, it would not take as long as the 11 hours it took to write it. This is because part of the hard work has already been done. As an IT worker spends time in an environment, he will little by little build a library of useful utilities and learn tricks to compensate for the defects of the existing software. He will also gain domain knowledge both in medical science and with the clinical aspect of the hospital environment. Because of this, his relative efficiency will grow as he has to spend less time writing code, and has a better understanding of his role within the medical research process.

This is what we hinted at earlier when we said that the value an IT worker provides goes beyond just freeing up some time for the medical personnel. As he works, the IT worker not only solves the problem at hand, but makes it easier to solve a similar problem in the future.

While we were working in the residents office in the hospital, another resident was busy with another data-entry task. While this task could not be fully automated because it required the resident to analyze a textual entry made by a doctor\footnote{Reliable natural language processing is an open problem in artificial intelligence.}, our program could be used to make it easier. We could have tweaked our program to fetch the textual entries and scan them for keywords, and then let the resident decide, using a well-thought user interface, what parts were relevant to her problem. This would have let the resident concentrate on the interesting problem and use her hard-gained medical skills while automating away the tedious, repetitive and error-prone bits (multiple clicks and keypresses to access one record). We estimate that we could have tweaked the program in about one hour, freeing between 5 and 20 hours for the resident, depending on the number of entries she had to process. This bumps our efficiency $\alpha$ between $5$ and $20$. Expecting an efficiency of $\alpha=10$ for an average IT worker after some time spent in the hospital looks like  a conservative estimate to us.

An IT worker can also discover ways to access data that was previously deemed too hard to gather to bother with it. Easy access to data means it is easier to do research. This in turn means better research, which gives better outreach, brings funds to the hospital and globally makes a better world for everyone. 

\subsection{Cost of opportunity}

In a world where there are infinitely many good doctors, nurses and residents, diverting one of these people's time to anything else than clinical activities or medical research costs only as much as this person's cost of employment.

We do not live in such a world. Residents don't count their hours, but still do not manage to work more
than 120 hours a week\footnote{Lazy bastards :/}. A resident-hour is a scarce resource, because there is not an infinite amount of resident that can be summoned and fired at will. Diverting a resident's time to a task like data-entry costs more than the resident's salary and some taxes. One must take into account the cost of opportunity of not using the residents where they shine : learning from doctors, teaching younger residents, trying to heal people and doing research.

Some of these tasks can be accomplished by other personnels (nurses or doctors depending on the task), but then the resident won't gain experience. They can be accomplished by stretching hours, but a tired medical practitioner is not as efficient as, and certainly more dangerous than a well-rested one. Or they can be foregone alltogether, which will degrade any useful metric of the hospital's utility.

Hiring an IT worker can be seen as hiring medical workers by proxy, at a very low cost. Hiring an IT worker with an efficiency of $\alpha = 3.5$ (our baseline, we expect $\alpha=10$) at $21$ €/h is equivalent to hiring $3.5$ more residents at $6$ €/h (i.e. less than their already ridiculously low cost).

% xkcd facile difficile

%Benefits are more (because network effect) for IT and costs for med are higher (because cost of opportunity).

%Network effect and scarcity of medical resident time (opportunity cost)

\section{Using our software}
\label{sec:deployment}

The source code of our software can be found at \url{https://gist.github.com/edouardklein/15b1895b64496267fc06}.

Our technical solution to the problem of diverting skilled personnel to dumb tasks can be used by anyone with a modicum of power.  An intern from any software or Computer Science (CS) cursus should be able to understand, copy and adapt this software quite quickly. If the analysis of \autoref{sec:comparative} has not convinced the reader of hiring a full fledged IT worker, or if the reader does not hold enough power to hire more than an intern, then starting with that to test the waters is an easy enough step to take. For follow-up research, we welcome data about the costs and benefits of actually going through with our advice.

If our reader was convinced by our arguments, or hired an intern and was happy with the result, we suggest  hiring, at an entry- to mid-level salary, a programmer with systems administration knowledge, a desire to learn (and self-teaching capabilities), a scholarly interest in the medical field, and a capacity to listen to, understand and adapt to what the users (in this case the professors, doctors, medical residents, midwives and nurses of the unit) say.

This person will be tasked with making the life of the hospital personnel as painless as possible with regards to IT. Provided he proved he can be trusted, he is to be let free to pursue what he assesses to be useful, and should only answer to one person, preferably the head of department, who will  refrain from ever micro-managing him, and will try his best to understand the nuts and bolts of the IT side of things.

One can not emphasize how important it is that the hired person is kind and understanding with the users, whom he was hired to serve, and that in return he is virtually free to set his own goals, deadlines and priorities.

A successful hire will bear fruits very quickly. Medical records will be created faster and with less mistakes, the time thus gained by the medical staff will be spend actually doing medicine or research, which will increase the quality of life of the patients and the throughput of the department, which always looks good during a department review. Little by little, the IT worker will stop working against the hindrances of the badly conceived hospital IT system, and will actually begin to create new, better solutions.

In an environment where most workers are highly skilled, one must not underestimate the cost of wasting a worker's time, and the benefits of making his life easier by hiring a complementary-skilled worker. There is a reason why doctors don't repair the coffee machine, do the electrical wiring or replace a broken window. Why should they have to tackle any IT or administrative tasks ?

\section{The big picture}

We just explained how, at the department level, one IT worker will provide far more value than what he costs. We propose here a set of decision to be taken at a higher level of power in the institution that we hope will allow for a generalize optimization of IT expenditures and will have many positive externalities.

We will first define what we want to happen, and who is susceptible to make those changes happen. We will then try to devise incentives that will push IT workers in the right direction, and warn the reader about "common-sense" policies that actually present perverse incentives that would have the opposite of the desired effect.

\subsection{Up to date software}
\label{sec:uptodate}

If the software that manages medical records had been well crafted, not even state-of-the-art or "advanced" by any measure, but just good and well-thought, then it would have, for example, presented us with a RESTful API \cite{fielding2000architectural} serving JSON. In other words, we could have pointed our program at an URL like \url{https://example.com/patients/123456789/record.json} and got all the medical records for patient number 123456789, in a format readable by humans yet easily parsable by a computer, the JSON format. Such a record may look like this :

\begin{lstlisting}[language=json,firstnumber=1]
"id":123456789,
"records":[
...
{
...
"Executing_MU":"Hematologie (LMR)",
"date":"2014-11-11T11:11:11.111Z",
"data":{
...
"Hemoglobine":10.6,
...
},
...
},
...
]
\end{lstlisting}

API and encoding formats should be documented, but one can see that a properly constructed API and format such as the ones we suggest here are more or less self documenting. Once we get a list of records such as these, finding what we need (the newest hemoglobin stat before the C-section, and the oldest after) is trivial and can be done in 5 lines of code or less.

Serving a request should take the server less than one second. Writing the code that send the request and parse the answer is less than an hour's job. Execution of the record-selection code is almost instantaneous (even if there are thousands of records). So, all in all, if we add the time it takes to write the program (less than an hour) and the time it takes to execute it (around one second per patient, so less than 20 minutes for 1100 patients), we get between one and two hours for $t_{IT}$. As we remember, the time it would have taken the resident was 41 hours. This puts the efficiency $\alpha$ of an IT worker dealing with good software somewhere between $20$ and $40$. This is highly desirable. So how do we get from the dreadful state of hospital IT we are in now to a state where we can truly begin to leverage technology's power ?

\subsection{Good solutions come from the bottom up}
\label{sec:good}

IT is a field job. Most problems happen locally with the users. They should be part of the solution. The best solutions are those that answer a real need, and evolve in response to changes in the eco-system (here the users' needs, the legal constraints, the medical advances etc.).

Therefore, an IT solution can only be good if it is deployed with boots on the ground. That is to say, with IT workers that can train the users, see the problems a submit corrections. Sadly, most of IT is organized the exact opposite way : we pay a high amount (in the order of millions) some external company to build a piece of software that barely answer the specifications, but that does not really matter because the specifications were defined by a central decision making body that was almost completely disconnected with the field. No one will ever ask the actual users what they needs are, or god forbid come and see them interact with the software to identify the pain points. The system will cost an arm to develop and the poor souls in the centralized (because centralized means more efficient, of course !) IT support team will have to maintain it for years to come, while nobody actually using it can say a word about what should be updated or re-thought.

Cristal-Net is not an isolated case of a centrally ordered failed software project. Louvois \cite[in french]{louvois}, the payroll management system for the army, is another appalling example of waste and mismanagement of public funds.

There is a need for a central influence, a driving force defining strategic objectives. But this does not mean a central body of decision makers giving away millions of euros for crap software.

One IT worker in a hospital unit can greatly increase, through tricks and ad-hoc solutions, the productivity of his immediate surroundings. A dozen or so IT workers, working in close collaboration with their medical colleagues in their respective department, but also collaborating as an IT group, exchanging ideas, code, software and information can bring any useful metrics of an hospital through the roof through the network effect. Once IT-workers are numerous and well integrated in whatever department they work in, consensus solutions will appear as if by magic. And they will share the qualities of modern open source software : relevance, reliability, security, and overall quality. In other words, low cost, high value. It is possible to manage the IT resources of a big state institution this way, as demonstrated those recent years by the French Gendarmerie (see \autoref{sec:related}).

\subsection{The right incentives}
\label{sec:incentives}

\paragraph{Intrinsic motivation}

In \autoref{sec:deployment}, we saw how to bootstrap the political solution by deploying our emergency  technical patch. Hiring an IT worker can be done at various levels of decision making (as an intern to a doctor, by a department head, by the hospital manager or at the central decision making level), and it will snowball from there. The decision is easy to make and good consequences will automatically follow (once again, we welcome data about implementing our advice).

Once the professors and doctors start hiring IT people in their department, and these IT people, thanks to the network effect, start creating new and better technical solutions to the various challenges of information management in hospitals, the time comes to take a more formal political approach. The goal is to make official and manage what may have started as different people at various level in the hierarchy following our advice of hiring an IT crew, and also to drive global optimization in order to harvest as much value as possible from the IT workers. Somewhat paradoxically, the best way to do that is to do as little as possible and let the knowledgeable people sort it out themselves. Collaboration and cooperation is deeply ingrained in IT and CS research culture\footnote{As things like open source software, open formats \cite{openformats} or the IETF \cite{ietf} prove.} and competent people will be able to work together.

We know from scientific studies \cite{pink2011drive} on motivation that for creative work, intrinsic motivation work best. Therefore, once the matter of salary is settled\footnote{The APHP is able to throw away millions of euros for big software projects that don't work, they can set aside some of it to pay in-house developers.}, and we saw \autoref{sec:economics} that the value provided by an IT worker will be higher than what a senior project manager costs to his employer, one should concentrate on fostering an atmosphere of creativity and collaboration.

\emph{Autonomy} is one of the tenants of intrinsic motivation, and we talked in \autoref{sec:deployment} of how an IT worker should answer only to the head of the department he works in. \emph{Purpose} should be familiar to anyone working in an hospital : people are not in it for the money. Hospital matters and the IT worker will be happy to help oil the cogs of a machine that tries its best to heal people. Working in close collaboration with medical personnel will yield rewards as one sees nurses and doctor becoming more efficient as the IT system gets better. \emph{Mastery} is the third tenant of intrinsic motivation. It consists in getting better at something that matters. This is where a central driving force can help. One should push the IT workers to hold conferences and workshops to discuss tips, tricks, technical advances etc, using \emph{peer-recognition} as an incentive for innovation and documentation.

Hiring people with a scholar background (such as CS PhDs) will also foster collaboration with the medical personnel in research activities. Being able to jump back into a research position after a technical bout may appeal to a lot of young graduates.

We also suggest that every decision impacting IT in hospitals should be vetted by a qualified majority of those IT workers. In other world, a opposition from small number of IT worker should be enough to cancel a project. This will avoid wasteful fiascos such as Cristal-Net and Louvois.

\paragraph{Kill the common sense} 

This system will auto-organize and work if no perverse incentives are present. Perverse incentives are well-intentioned mechanisms that will in fact be detrimental to the pursued goal. The role of the central decision makers will be, far more than taking action to help the IT workers, to avoid taking detrimental decisions such as trying to control and check everything.

The IT workers should have job security. Otherwise, they will avoid documenting the systems they create and voluntarily obfuscate code in order to become essential. Their pay, benefits and working conditions should be subject only to one person in the hierarchy (preferably the head of department), and not tied to any metric whatsoever. This would lead to pernicious optimization where people optimize the metric instead of doing their job. The same kind of problem appear when trying to assess scientific research output (see e.g. \cite{dora}). The problem is that bureaucrats with little to no training in the subject matter try to measure efficiency by imposing quantified goals (e.g. for IT "tickets solved per day", "number of lines of code written" or "number of hours spent in the office") the optimization of which is only loosely related to the actual goal.

The IT budget should not be subject to negative change over years. Savings from one year should be transferable to the following year. This will avoid end-of-year stupid purchases where one tries to burn the budget in order to get the same amount the year after. Equipment roll-out is expensive and usually happen in bursts. Only when a 10-year accumulation of funds is clearly unneeded a plan to give them back to the administration should be created under the IT worker's supervision.

IT workers at the department level are not only the missing link between decision makers and users, but should be an unavoidable part of the decision making process. No need to consult them, and create commissions, and hold meetings, and take minutes, and write reports. Just let them vote and veto any silly project that come from above. Good IT solutions come from the ground, organically.

\section{Related work}
\label{sec:related}

The agile trend in software development methodologies \cite{martin2003agile} put emphasis in continued and meaningful interaction between the developers and the end users. This is typically lacking in the big, costly and failed software projects that plague the IT budgets of state institutions.

\citet{lakhani2005hackers} studied the motivations of open source contributors and found that external motivators were not the primary driving force. On the contrary, intrinsic motivators such as creativity, fulfilling user's need and bettering one's programming skills were the primary incentives. This is in line with the incentives we propose to present hospital IT workers with.

The French Gendarmerie started converting their IT to open source in the early 2000's, with great success : IT costs reduction and better IT security (\cite{zdnet}, in French).

The \emph{Write The Docs} \cite{wtd} initiative regroups tools and methodologies to help write software documentation. One can forgive the volunteers writing open source software when they don't document their code. This is harder to do for a very expensive professionally developed solutions such as Cristal-Net.

\citet{menachemi2005hospital} authored a study of IT in Florida hospitals finding a positive relationship between IT use and efficiency. IT use is linked with higher expenses, though.

Another study by \citet{chaudhry2006systematic} surveys statistics about IT use in hospitals. The study is broader than what we are talking about here, but one take-away with regards to IT efficiency in hospitals is that data is lacking.  We reproduce part of their conclusion here :
\begin{quote}
This review suggests several important future directions in the field. First, additional studies need to evaluate commercially developed systems in community settings, and additional funding for such work may be needed. Second, more information is needed regarding the organizational change, workflow redesign, human factors, and project management issues involved with realizing benefits from health information technology. Third, a high priority must be the development of uniform standards for the reporting of research on implementation of health information technology, similar to the Consolidated Standards of Reporting Trials (CONSORT) statements for randomized, controlled trials and the Quality of Reporting of Meta-analyses (QUORUM) statement for meta-analyses (109, 110). Finally, additional work is needed on interoperability and consumer health technologies, such as the personal health record.
\end{quote}
We believe our suggested approach would help address those issues, especially the second and last ones.

\section{Conclusion}

Appalled by the dreadful state of IT in Paris hospitals, and ashamed of the wasteful IT policies that spend millions of euros to end up in such a sad state, we propose mid- to long-term low-budget provisions that will allow for :
\begin{itemize}
\item a decrease, down to complete removal, of loss of skilled manpower to data-entry tasks thanks to the use of up-to-date technology (\autoref{sec:uptodate});
\item a decrease, down to near-optimal levels, of software creation, deployment and maintenance cost, due to the hiring of people skilled in IT, immersed in the day-today life of a hospital unit and with a say in the decision process of all things IT-related (\autoref{sec:incentives});
\item an increase in medical research productivity and scope due to the complete availability of rich, ever growing, structured datasets, and the freeing of precious medically skilled man-hours, that could be reinvested in research or clinical endeavours (\autoref{sec:compounding}).
\end{itemize}

Based on our experience of writing a piece of software to save 41 hours of work to one medical resident, we provide one data point of quantitative empirical evidence backing up our claims.

It is our hope that our advice will be followed by some heads of department, which will allow for data gathering about the efficiency of IT workers and will either give weight to our arguments, or give us an opportunity to revise our proposal. We are quite confident that our advice is good, though, as the situation is so bad that to make it worse one would have to actively try.

\printbibliography
\end{document}